\documentclass[preprint,showpacs,prb]{revtex4}
\usepackage{amssymb}
\usepackage[dvipdfm]{graphicx}

\begin{document}

\title{Roughness-induced energetic disorder at the metal/organic interface}
\author{S.V. Novikov}
\affiliation{A.N. Frumkin Institute of Physical Chemistry and
Electrochemistry, Moscow 119991, Russia}
\author{G.G. Malliaras}
\affiliation{Materials Science and Engineering, Cornell
University, Ithaca, NY 14853-1501}


\begin{abstract}
The amplitude of the roughness-induced energetic disorder at the
metal/organic interface is calculated. It was found that for
moderately rough electrodes, the correction to the electrostatic
image potential at the charge location is small. For this reason,
roughness-induced energetic disorder cannot noticeably affect
charge carrier injection, contrary to the recent reports.
\end{abstract}

\maketitle

Effective injection of charge carriers is a major requirement for
efficient and reliable performance of electronic organic devices
(light emitting diodes, thin film transistors, and
others).\cite{e3} Understanding charge injection is intimately
connected to the structure of the metal/organic interface (or the
interface of an organic material with another conductive material,
such as indium tin oxide), and the detailed knowledge of this
structure is important for other processes and applications.
Recent experimental studies indicate that the Richardson-Schottky
thermionic injection
\begin{equation}
j \propto \exp\left(-\frac{e\phi}{kT}+\gamma\sqrt{E}\right)
 \label{RS}
\end{equation}
is a good starting point for the description of injection process
in organic devices.\cite{e1,e2,e3} Here $j$ is the injected
current density, $\phi$ is the height of the barrier at the
interface, and $E$ is an applied electric field. At the same time,
measurements at low temperature show that the decrease of the
injection current density is much smaller than the anticipated
decrease according to the Richarson-Schottky model.\cite{t1,t2} It
was suggested that the reason for this discrepancy is the effect
of energetic disorder in the organic material.\cite{e1} Usually,
the calculation of the effect of disorder on injection is carried
out using disorder parameters estimated from the charge transport
data;\cite{t1,t2} this means that these parameters describe the
disorder in the bulk of the organic material. The experimental
data clearly indicates, though, that in some cases, a surface
dipolar layer is formed at the metal-organic
interface.\cite{ishii,cahen} It is very reasonably to assume that
this layer has some degree of disorder and, hence, will provide an
additional contribution to the energetic disorder at the
interface.\cite{baldo1,baldo2}

Recently, roughness at the metal/organic interface was suggested
as the source of additional energetic disorder, localized near the
interface.\cite{baldo2} A calculation of the standard deviation of
the disorder $\sigma(z_0)$ for a point charge $e$ located at
distance $z_0$ from the mean plane of the weakly rough metal
surface having profile $h(x,y)$ may be carried out in the
following way (we assume that the mean plane of the electrode is
located at $z=0$). Let us suppose that $z_0 \ll l$, where $l$ is
the surface correlation length. Then we can consider the electrode
surface at the vicinity of a charge as a flat plane and treat
surface deviation $h(x,y)$ from the mean plane as a constant. The
change of the image potential at the charge location due to the
shift of the surface position by $h$ is (in the first order in
$h$)
\begin{equation}
\delta \varphi(z_0)\approx -\frac{eh}{2\varepsilon z_0^2}
\label{first}
\end{equation}
and $\sigma(z_0)$ is estimated as
\begin{equation}
\sigma^2(z_0)=e^2 \left<\left[\delta \varphi(z_0)\right]^2\right>=
\frac{e^4h_0^2}{4\varepsilon^2 z_0^4}, \label{baldo_sigma}
\end{equation}
where the angular brackets denote an average over the ensemble of
realizations of the surface roughness,
$h_0^2=\left<h^2(x,y)\right>$ is the roughness variance and
$\varepsilon$ is a dielectric constant. The mean plane of the
electrode is defined in such a way that $\left<h(x,y)\right>=0$.
Eq. (\ref{baldo_sigma}) is exactly the result of
Ref.~\onlinecite{baldo2}, though obtained in a much simpler way.

\begin{figure}
\begin{center}
\includegraphics[width=2.5in]{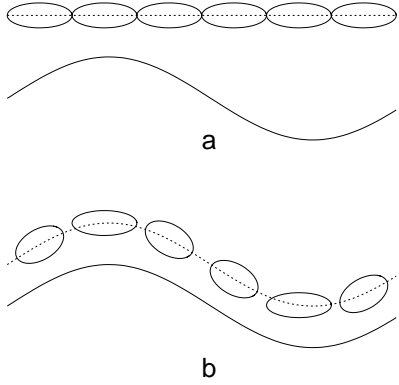}
\end{center}
\caption{Two models of the arrangement of organic molecules
 (ellipsoids) at the rough
surface of the electrode (solid line): a) first layer is located
at constant distance to the mean electrode plane;\cite{baldo2} b)
molecules in the first layer are located at constant distance to
the surface of the electrode.} \label{fig1}
\end{figure}

The important parameter for charge injection is the energetic
disorder directly at the interface, i.e., in first several layers
of organic transport molecules adjacent to the electrode. If for
the very first layer $z_0=6$ \AA, $h_0=3.5$ \AA, and
$\varepsilon=1$, then $\sigma(z_0)=0.7$ eV
(Ref.~\onlinecite{baldo2}). Yet the validity of Eq.
(\ref{baldo_sigma}) for short distances is very dubious because of
the basic assumption that organic molecules in any particular
layer of organic material are situated at a \textit{constant
distance $z_0$ from the mean plane of the electrode}\cite{baldo2}
(see Fig. \ref{fig1}a). The size of a typical transport molecule
(8-10 \AA) is small in comparison to the surface correlation
length (typically, $l=20-50$ nm,  Ref.~\onlinecite{ITO}). In this
situation it is natural to expect that a better model of the
interface is one where the molecules follow the electrode profile,
and any particular layer is located at the \textit{constant
distance $z_0$ to the actual surface of the electrode} (Fig.
\ref{fig1}b). We will see that in this model $\sigma(z_0)$ differs
drastically from the estimation in Eq. (\ref{baldo_sigma}). The
calculation of $\sigma(z_0)$ along the profile of the rough metal
surface is the major difference between our paper and the paper of
Rahman and Maradudin,\cite{maradudin} where the general expression
for the mean image potential for a rough dielectric interface was
obtained.

The potential for a point charge located at the vicinity of a
rough metal surface obeys the Poisson equation
\begin{equation}
\Delta \varphi=-\frac{4\pi
e}{\varepsilon}\hskip2pt\delta\left(\vec{r}-\vec{r}_0\right)
\label{Poisson}
\end{equation}
with  the boundary condition
\begin{equation}
\left.\varphi\right|_{z=h(x,y)}=0 \label{bound1}
\end{equation}
We assume that the roughness is small $h_0/l \ll 1$, and Eq.
(\ref{Poisson}) can be treated via a perturbation theory approach.
We are going to calculate the leading contribution only. A
possible approach to perform this calculation is to transform to
coordinates $z_{\textrm{new}}= z-h(x,y)$, so for new $z$ the
boundary condition is set for $z=0$. In the new coordinates, the
Poisson equation takes the form
\begin{eqnarray}
&\Delta_{\bot}\varphi+\frac{\partial^2 \varphi}{\partial
z^2}\left[\left(\frac{\partial h}{\partial
x}\right)^2+\left(\frac{\partial h}{\partial y}\right)^2+1\right]-
\label{Poisson2}\\* &2\left(\frac{\partial^2 \varphi}{\partial x
\partial z } \frac{\partial h}{\partial x }+\frac{\partial^2 \varphi}{\partial
y
\partial z } \frac{\partial h}{\partial y }\right)
-\frac{\partial \varphi}{\partial z } \Delta_{\bot}h=-\frac{4\pi
e}{\varepsilon}\hskip2pt\delta\left(\vec{r}-\vec{z}_0\right),
\nonumber
\end{eqnarray}
where $\Delta_{\bot}$ is a two-dimensional (2D) Laplacian and we
assume $\vec{z}_0=(0,0,z_0)$. Note that in the new coordinates the
condition  $z_0=\textrm{const}$ is approximately equivalent to a
constant distance to the profile of the electrode (with a small
correction proportional to $h^2_0/l^2$ and insignificant to our
analysis).

Let us try to find a formal solution as a series
\begin{eqnarray}
\varphi=\sum_{n=0}^{\infty} \varphi_n, \hskip10pt \varphi_n
\thicksim O(h^n), \hskip10pt \left.\varphi_n\right|_{z=0}=0,
\label{perturb}\\*
\varphi_0(\vec{r})=\frac{e}{\varepsilon\left|\vec{r}-\vec{z}_0\right|}
-\frac{e}{\varepsilon\left|\vec{r}+\vec{z}_0\right|}.\nonumber
\end{eqnarray}
The first-order correction is
\begin{equation}
\varphi_1(\vec{r})=\int d\vec{r}_1
G(\vec{r},\vec{r}_1)J(\vec{r}_1), \label{phi_1}
\end{equation}
here $G(\vec{r},\vec{r}_1)$ is the Green function for the Laplace
operator with zero boundary condition at $z=0$, while the source
term is
\begin{eqnarray}
&J(\vec{r})=2\left(\frac{\partial^2 \varphi_0}{\partial x
\partial z } \frac{\partial h}{\partial x }+\frac{\partial^2 \varphi_0}{\partial
y
\partial z } \frac{\partial h}{\partial y }\right)+\frac{\partial \varphi_0}{\partial z }
\Delta_{\bot}h=\nonumber\\*
&-\frac{e}{\varepsilon}\frac{\partial}{\partial
z_0}\left[2\left(\frac{\partial P}{\partial x}\frac{\partial
h}{\partial x }+\frac{\partial P}{\partial y}\frac{\partial
h}{\partial y }\right)+P\Delta_{\bot}h\right], \label{J}\\*
&P(\vec{r})=\frac{1}{\left|\vec{r}-\vec{z}_0\right|}
+\frac{1}{\left|\vec{r}+\vec{z}_0\right|}.\nonumber
\end{eqnarray}
Note that, in our case, the correction to $\varphi_0$ depends not
on $h(x,y)$ itself, but on its derivatives and vanishes for
$h(x,y)=\textrm{const}$, as it should be.

\begin{figure}
\begin{center}
\includegraphics[width=3in]{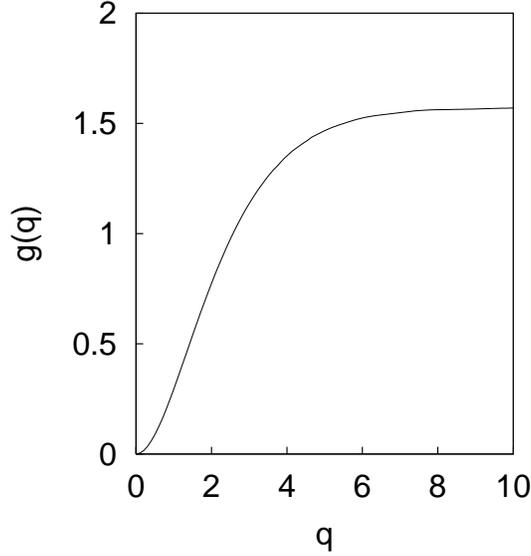}
\end{center}
\caption{Function $g(q)$.} \label{fig2}
\end{figure}

The Green function has the form\cite{synthmet}
\begin{equation}
G(\vec{r},\vec{r}_1)=\frac{1}{4\pi^2}\int d\vec{k}
e^{-i\vec{k}\left(\vec{\rho}-\vec{\rho}_1\right)}G_k(z,z_1),
\label{Green_2}
\end{equation}
where $\vec{k}$ and $\vec{\rho}=(x,y)$ are 2D vectors, and the
 Green function $G_k(z,z_1)$ obeys the equation
\begin{equation}
\frac{d^2G_k}{dz^2}-k^2G_k=\delta(z-z_1), \hskip10pt G_k(0,z_1)=0.
\label{Green_Z}
\end{equation}
The solution of Eq. (\ref{Green_Z}) is
\begin{eqnarray}
G_k(z,z_1)=-\frac{1}{k}\sinh kz_{-}\exp(-kz_{+}),\\ z_{+}={\rm
max}(z,z_1), \hskip10pt z_{-}={\rm min}(z,z_1).\nonumber
\end{eqnarray}
We are going to calculate the correction (\ref{phi_1}) for
$\vec{r}=\vec{z}_0$ only, because
$\sigma^2(z_0)=e^2\left<\varphi_1^2(\vec{z}_0)\right>$. A simple
but lengthy calculation gives for $\varphi_1$
\begin{widetext}
\begin{equation}
\varphi_1\left(\vec{z}_0\right)=\frac{e}{4\pi^3\varepsilon}\int
d\vec{k}_1 d\vec{k}_2 h_{\vec{k}_1-\vec{k}_2}
\exp\left[-(k_1+k_2)z_0\right]\left[\cosh\left(k_1-k_2\right)z_0-1\right],
\label{S_1}
\end{equation}
\end{widetext}
and $h_{\vec{k}}$ is a Fourier transform of $h(\vec{\rho})$. The
integral (\ref{S_1}) can be simplified further. Let us make a
transition to new vector coordinates
$\vec{p}=(\vec{k}_1+\vec{k}_2)/2$, $\vec{q}=\vec{k}_1-\vec{k}_2$.
Then
\begin{equation}
\varphi_1(\vec{z_0})=\frac{e}{4\pi^3\varepsilon z_0^2}\int
d\vec{q} g(qz_0) h_{\vec{q}}, \label{S_1a}
\end{equation}
where
\begin{eqnarray}
&g(q)=\int d\vec{p} \hskip2pt \exp\left(-R\right) \left(\cosh Q
-1\right), \label{g(q)}\\
&R=\left|\vec{p}+\frac{1}{2}\vec{q}\right|+\left|\vec{p}-\frac{1}{2}\vec{q}\right|,
\hskip10pt
Q=\left|\vec{p}+\frac{1}{2}\vec{q}\right|-\left|\vec{p}-\frac{1}{2}\vec{q}\right|.\nonumber
\end{eqnarray}
Function $g(q)$ can be easily estimated for $q \ll 1$ and $q \gg
1$. In the first case we can expand the hyperbolic cosine in  the
integral (\ref{g(q)}) in power series of $\vec{q}$, thus obtaining
\begin{equation}
g(q)\approx\int d\vec{p} \hskip2pt
\left(\frac{\vec{p}\cdot\vec{q}}{p}\right)^2\exp\left(-2p\right)=
\frac{\pi q^2}{8}. \label{small_q)}
\end{equation}
In the opposite case ($q \gg 1$) analysis shows that the only
significant (and equal) contributions to Eq. (\ref{g(q)}) goes
from $\vec{p}\approx\pm\frac{1}{2}\hskip2pt\vec{q}$, so setting
$\vec{p}=\frac{1}{2}\hskip2pt\vec{q}+\vec{s}$ we have
\begin{equation}
g(q)\approx \int d\vec{s} \hskip2pt\exp\left(-2s\right)
=\frac{\pi}{2}.
 \label{large_q)}
\end{equation}
The general behavior of $g(q)$ is shown in Fig. \ref{fig2}.

 Finally, the variance of
roughness-induced energetic disorder is
\begin{equation}
\sigma^2(z_0)= e^2\left<\varphi_1^2(\vec{z_0})\right> =\frac{e^4
h_0^2}{4\pi^4\varepsilon^2 z_0^4}\int d\vec{q}
g^2(qz_0)C_{\vec{q}}. \label{final}
\end{equation}
This equation is the major result of this paper. Here
$C_{\vec{q}}$ is the Fourier transform of the surface correlation
function, which we define in a usual way (assuming spatially
homogeneous roughness)
\begin{equation}
\left<h(\vec{\rho})h(\vec{\rho}_1)\right>=
h_0^2C(\vec{\rho}-\vec{\rho}_1)  \label{corr}
\end{equation}
with $C(0)=1$, so $\left<h^2(\vec{\rho})\right>=h_0^2$. For
homogeneous roughness
\begin{equation}
\left<h_{\vec{k}}h_{\vec{k}_1}\right>=4\pi^2 h_0^2 C_{\vec{k}}
\hskip2pt\delta\left(\vec{k}+\vec{k}_1\right). \label{Ck}
\end{equation}
If $z_0 \gg l$, then we can replace $g(qz_0)$ by its limit value
of $\pi/2$, and in this case
\begin{equation}
\sigma^2(z_0) \approx \frac{e^4 h_0^2}{4\varepsilon^2 z_0^4}.
\label{large_z0}
\end{equation}
This result is equivalent to  Eq. (\ref{baldo_sigma}), but it is
valid only far away from the rough electrode surface. The reason
for the equivalence of Eq. (\ref{large_z0}) and Eq.
(\ref{baldo_sigma}) is the need to cancel the leading term in Eq.
(\ref{S_1a}) in the old (physical) coordinate system. Indeed, as
it follows from Eq. (\ref{S_1a}) for $z_0 \gg l$
\begin{equation}
\varphi_1\left(z_0\right)=\frac{eh}{2\varepsilon
z_0^2}+o\left(\frac{1}{z_0^2}\right)\Omega\left[h\right],\label{oh}
\end{equation}
where $\Omega$ is some integral operator. Image potential at the
charge location is
\begin{eqnarray}
-\frac{e}{2\varepsilon z_0}+\frac{eh}{2\varepsilon
z_0^2}+o\left(\frac{1}{z_0^2}\right)\Omega\left[h\right] =
\label{cancel}\\ -\frac{e}{2\varepsilon
z_0^{\textrm{old}}}+o\left(\frac{1}{\left(z_0^{\textrm{old}}\right)^2}\right)\Omega\left[h\right]
+O(h^2)\nonumber
\end{eqnarray}
(here $z_0^{\textrm{old}}$ denotes the distance to the mean plane
of the electrode). This result means that in the old (physical)
coordinate system the correction to the image potential in the
first order in $h$ decays faster than
$1/\left(z_0^{\textrm{old}}\right)^2$ for large distances. This is
not surprising because this kind of decay is possible only for
$h\approx\textrm{const}$, which is not the case for $z_0 \gg l$,
where many uncorrelated domains of the rough surface contribute to
the image potential.

All these intricacies are not important for charge injection,
where a relevant distance to the  surface of the electrode is
small. If $z_0 \ll l$,  then
\begin{equation}
\sigma^2(z_0) \approx \frac{e^4 h_0^2}{256\pi^2\varepsilon^2 }\int
d\vec{q} q^4 C_{\vec{q}}\propto \frac{e^4 h_0^2}{\varepsilon^2
l^4}. \label{small_z0}
\end{equation}
The later estimation is valid if $C(\vec{\rho})$ can be
characterized by the scale $l$ only, and the integral in Eq.
(\ref{small_z0}) converges for $q\rightarrow \infty$.  If we
assume a Gaussian correlation function
\begin{equation}
C(\vec{\rho})=\exp\left(-\frac{\rho^2}{2l^2}\right) \label{GC}
\end{equation}
which is a good approximation for indium tin oxide
electrodes\cite{ITO}, then for $z_0\ll l$
\begin{equation}
\sigma^2(z_0) \approx \frac{e^4 h_0^2}{8\varepsilon^2 l^4}.
\label{small_z0_G}
\end{equation}
Eq. (\ref{small_z0_G}) is similar to Eq. (\ref{baldo_sigma}) with
the only crucial difference: $z_0$ is replaced by $l$. For the
roughest electrode, mentioned in Ref.~\onlinecite{ITO}, with
$h_0=4$ nm and $l=14$ nm, we have at the interface $\sigma \approx
0.01$ eV.

Let us consider the case, when the integral (\ref{small_z0}) does
not converge for $q\rightarrow \infty$. This is the case of the
fractal rough surface with a correlation function
\begin{equation}
C_{\vec{k}}=\frac{Al^2}{\left(1+k^2 l^2\right)^{1+\alpha}},
\hskip10pt A=4\pi\alpha\left[1-\frac{1}{\left(1+k_c^2 l^2
\right)^\alpha}\right]^{-1}, \label{Self-A}
\end{equation}
here $0\le \alpha < 1$ (Ref.~\onlinecite{self}). In fact, any
fractal surface can be realized as an intermediate asymptotic
only, for some spatial scale range, thus the proper cut-off $k_c$
is assumed in Eq. (\ref{Self-A}). For a clear physical reason
(discrete nature of a real metal surface) $k_c \ll 1/a$ where $a$
is a typical interatom distance, while $k_cl \gg 1$. This means
that $z_0 k_c \lesssim 1$ for the organic layers closest to the
metal surface. Hence, we can still use the small-q asymptotic of
$g(q)$ and
\begin{eqnarray}
&\sigma^2(z_0)\approx \frac{A e^4 h_0^2 l^2}{128\pi\varepsilon^2
}\int_0^{k_c} dq \frac{q^5}{\left(1+q^2
l^2\right)^{1+\alpha}}\approx \label{Self-A-sigma} \\ &\frac{A e^4
h_0^2 \left(k_c l
\right)^{2(2-\alpha)}}{256(2-\alpha)\pi\varepsilon^2
 l^4}.\nonumber
\end{eqnarray}
In the most favorable for large $\sigma$ case $\alpha \approx 0$
\begin{equation}
\sigma(z_0)\approx \frac{\pi e^2 h_0 k_c^2}{8\varepsilon \sqrt{\ln
(k_c l)}}
 \label{a=0}
\end{equation}
and for $h_0=5$ {\AA}  the value of $\sigma$ becomes comparable
with the bulk value of 0.1 eV only for $k_c \gtrsim 0.1$
\AA$^{-1}$. Such value for $k_c$ seems to be unreasonably large.
Scanning microscopy data indicate that typically the Gaussian
correlation function (\ref{GC}) is a good approximation for rough
electrode surfaces at nanometer scale.\cite{ITO} Fractal surfaces
have been indeed observed in clusters formed by small metal
particles, but the relevant spatial scale was very different: even
the size of the individual metal particle was no less than 10 nm
(Refs.~\onlinecite{fractal1,fractal2}); in this case $k_c < 0.01$
\AA$^{-1}$.

In conclusion, we have found that the contribution from image
forces to the roughness-induced energetic disorder is typically
too weak to provide a noticeable effect on injection, contrary to
the analysis in Ref.~\onlinecite{baldo2}.

This work was supported by the ISTC grant 2207 and RFBR grants
02-03-33052 and 03-03-33067. The research described in this
publication was made possible in part by Award No. RE2-2524-MO-03
of the U.S. Civilian Research \& Development Foundation for the
Independent States of the Former Soviet Union (CRDF).

\end{document}